\documentclass[a4paper]{article}
\usepackage{cite}
\usepackage{amsmath}
\usepackage{amsthm}
\usepackage{amssymb}
\usepackage{amsfonts}
\usepackage[latin1]{inputenc}

\theoremstyle{theorem}
\newtheorem{theorem}{Theorem}
\newtheorem{proposition}[theorem]{Proposition}

\theoremstyle{definition}
\newtheorem{definition}[theorem]{Definition}

\theoremstyle{remark}
\newtheorem{remark}[theorem]{Remark}

\newcommand{\al}{\alpha}

\newcommand{\ep}{\epsilon}

\newcommand{\ga}{\gamma}
\newcommand{\ka}{\kappa}

\newcommand{\om}{\omega}

\newcommand{\vp}{\varphi}

\newcommand\Om\Omega
\newcommand\Te\Theta
\newcommand{\bq}{\mathbf{q}}

\newcommand{\bp}{\mathbf{p}}

\newcommand{\bL}{\mathbf{L}}

\def\RR{\mathbb{R}}

\newcommand{\cH}{{\mathcal H}}

\newcommand{\cM}{{\mathcal M}}

\newcommand{\fsl}{\mathfrak{sl}}

\newcommand{\pd}{\partial}
\newcommand\minus\backslash

\newcommand{\ms}{\mspace{1mu}}
\newcommand\lan\langle
\newcommand\ran\rangle

\newcommand{\dd}{{\mathrm d}}
\newcommand{\sub}[1]{_{\mathrm{#1}}}

\newcommand\dq{\dot q}
\DeclareMathOperator\SO{SO} \DeclareMathOperator\U{U}

\newcommand{\Kep}{_{\mathrm{Kepler}}}
\newcommand{\Harm}{_{\mathrm{Harm}}}

\title{\Large A maximally superintegrable system on an $n$-dimensional space of nonconstant curvature}
\author{\normalsize\'A. Ballesteros$^a$\thanks{angelb@ubu.es} \and \normalsize
A. Enciso$^b$\thanks{aenciso@fis.ucm.es} \and\normalsize F. J.
Herranz$^a$\thanks{fjherranz@ubu.es} \and\normalsize O.
Ragnisco$^c$\thanks{ragnisco@fis.uniroma3.it}}
\date{\small $^a$ Depto.\ de F\'\i sica, Universidad de Burgos,
09001 Burgos, Spain\vspace{1ex}\\
$^b$ Depto.~de F{\'\i}sica Te\'orica II, 
Universidad Complutense, 
28040 Madrid, Spain\vspace{1ex}\\
$^c$ Dip.\ di Fisica, Universit\`a di Roma 3, and Instituto
Nazionale di Fisica Nucleare
\\Via Vasca Navale 84, 00146 Rome, Italy}

\begin{document}
\maketitle

\begin{abstract}
A novel Hamiltonian system in $n$ dimensions which
admits the maximal number $2n-1$ of functionally independent,
quadratic first integrals is presented. This system turns out to be
the first example of a maximally superintegrable Hamiltonian on an
$n$-dimensional Riemannian space of nonconstant curvature, and it can be
interpreted as the intrinsic Smorodinsky--Winternitz  system on such
a space. Moreover, we provide three different complete sets of integrals
in involution and solve the equations of motion in closed form.
\end{abstract}

\section{Introduction}

A Hamiltonian system on a $2n$-dimensional phase space is
\emph{maximally superintegrable} (MS) if it is Liouville
integrable~\cite{AKN97} and there exist further $n-1$ functionally
independent (global) first integrals. Throughout this paper we shall
restrict ourselves to the case of \emph{natural} Hamiltonians, whose
kinetic term defines a Riemannian metric on an $n$-dimensional
configuration space.

The intrinsic relevance of MS systems has resulted in a wealth of
results concerning MS Hamiltonians~\cite{Ra97,GP02,SIGMA,CRS06,KWMP02}. MS
models with integrals of motion quadratic in the momenta are of
particular interest because of their connections with generalized
symmetries~\cite{GLZ92,STW01}, isochronic
potentials~\cite{Ca04,Go04}, and separability of the associated
Hamilton--Jacobi (HJ) equation and Schr\"odinger
equations~\cite{Hu76,KM84,BMS01}. A systematic study of MS systems
on certain low-dimensional spaces is currently being developed by
Kalnins et al. (cf.~\cite{KKM06} and references therein), and in
fact in two dimensions they managed to obtain a (local)
classification of all MS Hamiltonians with integrals at most
quadratic in the momenta~\cite{KKMW02,KKMW03}. This result, which
hinges on a classical theorem by Koenigs~\cite{Ko72}, yielded the
first examples of MS systems not defined on constant curvature
spaces.

Unfortunately, most examples of MS Hamiltonians are restricted to
low-dimensional spaces. To the best of our knowledge, the only known
examples of $n$-dimensional MS systems with quadratic integrals are
given by the (generalized) Kepler problem~\cite{Ev90}, the
Smorodinsky--Winternitz (SW) system~\cite{FMSUW65,Ev90b}, their
generalizations to the simply connected spaces of constant sectional
curvature (cf.~\cite{BHSS03} and references therein), and the
geodesic flow on these spaces. Note that the allowed potentials
correspond essentially to those appearing in the extension to space
forms of Bertrand's theorem~\cite{Be1873,Li1923}. If one drops the
requirement that the integrals be quadratic in momenta, the list can
be enlarged with the rational and hyperbolic
Calogero--Sutherland--Moser models of type
$A$~\cite{Ca71,Su71,Mo75,Wo83}, the nonisotropic oscillator with
rational frequencies and the nonperiodic Toda
lattice~\cite{To67,ADS06}.

In this paper we show that the Hamiltonian
\begin{equation}\label{cH}
\cH(\bp,\bq)=\frac{\bp^2+\om^2\,\bq^2+\sum_j b_j\ms
q_j^{-2}}{\ka+\bq^2}\,,
\end{equation}
where $\ka>0$ and $\om^2,b_j\geq0$, is a MS system with quadratic
first integrals. Here $\bp,\bq\in\RR^n$, and we shall always assume
that $\RR^{2n}$ is endowed with its standard symplectic form
$\dd\bq\wedge \dd\bp$. For any $\mathbf u\in\RR^n$ we use the
notation
\[
\mathbf u^2=\mathbf u\cdot\mathbf u=\sum_i u_i^2
\]
and all the sum indices
run from 1 to $n$ unless otherwise
stated.

The Hamiltonian~\eqref{cH} yields the
motion under the potential
\begin{equation}\label{V}
V(\bq)=\frac{\om^2\,\bq^2+\sum_j b_j\ms q_j^{-2}}{\ka+\bq^2}
\end{equation}
on the conformally flat Riemannian manifold $\cM^n:=(\RR^n,\dd s^2)$
whose metric is given by
\begin{equation}\label{ds2}
\dd s^2=(\ka+\bq^2)\,\dd\bq^2\,.
\end{equation}
The Hamiltonian~\eqref{cH} provides the first example of a
Hamiltonian system on a Riemannian space of nonconstant curvature
which is MS in any dimension. The analysis of the geometry of
$\cM^n$ presented in Sec.~\ref{S:geometry} shows that this model can
be in fact regarded as the intrinsic SW system on this space. When
$n=2$ this model is listed in Kalnins et al.'s classification of MS
systems in the Darboux space of type III~\cite{KKMW03}. This model
is singled out among the others in this list because it is the only
one readily amenable to the Poisson coalgebra treatment developed in
Sec.~\ref{S:int}.

The paper is organized as follows. In Sec.~\ref{S:int} we show that
the Hamiltonian~\eqref{cH} is indeed MS and briefly study the
separability of the HJ equation. The proof is greatly simplified by
the use
 of the $\fsl(2)$ Poisson coalgebra
symmetry~\cite{BR98,BMR02} of the model. In Sec.~\ref{S:eqs} we
study the equations of motion and find their general solution in
closed form using the previously computed first integrals. Quite
remarkably, the radial motion is essentially that of the Kepler
problem, whereas the HJ equation essentially coincides with the one
for the SW system. In Sec.~\ref{S:geometry} we analyze the geometry
of the underlying Riemannian space and relate the
Hamiltonian~\eqref{cH} to the intrinsic harmonic oscillator in
$\cM^n$.

\section{Maximal superintegrability and separability of the HJ equation}
\label{S:int}

Instead of dealing with the model~\eqref{cH} directly, in the
following two sections we prefer to analyze the Hamiltonian
\begin{equation}\label{H}
H(\bp,\bq)=\frac{\bp^2-c+\sum_j b_jq_j^{-2}}{2(\ka+\bq^2)}\,,\qquad
c:=\ka\om^2\,,
\end{equation}
which is related to~\eqref{cH} via $\cH=2H+\om^2$. Obviously the
maximal superintegrability of $H$ is equivalent to that of $\cH$,
whereas the trajectories of $H$ can be mapped into those of $\cH$
via the time dilation $t\mapsto t/2$. In particular, $H$ and $\cH$
possess the same orbits.

A first observation is that there is an $\fsl(2)$ Poisson coalgebra symmetry of
the Hamiltonian~\eqref{H} that immediately provides $2n-2$ quadratic integrals.
This underlying coalgebra structure is shared by all the other known examples
of quadratically MS models~\cite{BH07,BHSS03}, as well as by uncountably many
other integrable Hamiltonian systems which are not MS. Therefore, the crucial
result in this section is the derivation of an additional  first integral,
functionally independent of the previous $2n-2$.

The $\fsl(2)$ Poisson coalgebra is defined by the basis
$\{J_\ep:\ep=\pm,0\}$ together with the following Lie--Poisson
brackets, Casimir $C$ and primitive coproduct $\Delta$:
\begin{gather}
\{J_0,J_+\}=2J_+\,,\qquad \{J_0,J_-\}=-2J_-\,,\qquad
\{J_-,J_+\}=4J_0\,,\label{brackets}\\
C=J_-J_+-J_0^2\,,\qquad \Delta(J_\ep)=J_\ep\otimes1+1\otimes
J_\ep\,.\label{Casimir}
\end{gather}
A symplectic (one-dimensional) realization of the
algebra~\eqref{brackets} on $(\RR^2,\dd q_1\wedge \dd p_1)$ is given
by
\[
J_-=q_1^2\,,\qquad J_0=q_1p_1\,,\qquad J_+=p_1^2+b_1q_1^{-2}\,,
\]
where $b_1$ is a real parameter labeling the representation which corresponds
to the value of the (one-dimensional) Casimir. Under this realization, the
Lie-Poisson product of the $\fsl(2)$ algebra is recovered by computing the
corresponding Poisson bracket in $(\RR^2,\dd q_1\wedge \dd p_1)$. Moreover, its
$n$-th coproduct yields a symplectic realization in $\RR^{2n}$ via
\begin{equation}
J_-=\bq^2\,,\qquad J_0=\bp\cdot\bq\,,\qquad J_+=\bp^2+\sum_jb_j\ms
q_j^{-2}\,,\label{Jep}
\end{equation}
with $b_j\in\RR$. Again, the Lie-Poisson product is given by the standard
Poisson bracket
\[
\{f,g\}=\frac{\pd f}{\pd\bq}\frac{\pd g}{\pd\bp}-\frac{\pd
f}{\pd\bp}\frac{\pd g}{\pd\bq}
\]
and the Casimir~\eqref{Casimir} is simply
\[
C=\bL^2+\sum_j\frac{b_j\ms\bq^2}{q_j^2}\,,
\]
where $\bL^2$ denotes the angular momentum. From this expression it
is apparent that $C$ is a homogenous function of degree 0.

This $\fsl(2)\otimes\cdots\otimes\fsl(2)$ symmetry endows any
$n$-dimensional Hamiltonian in the enveloping algebra with $2n-3$
integrals other than the Hamiltonian given by the left and right
partial Casimirs. More precisely we have the following result, which
we quote from Ref.~\cite{BH07}.

\begin{theorem}\label{T:2n-3}
Any Hamiltonian $H_0(\bp,\bq)=h(J_+,J_0,J_-)$ possesses $2n-3$ first
integrals given by the left and right partial Casimirs
\begin{align*}
C^{(m)}&=\sum_{1\leq i<j\leq
m}\bigg[(q_ip_j-q_jp_i)^2+\frac{b_iq_j^2}{q_i^2}
+\frac{b_jq_i^2}{q_j^2}\bigg]+\sum_{i=1}^mb_i\,,\\
C_{(m)}&=\sum_{n-m<i<j\leq
n}\bigg[(q_ip_j-q_jp_i)^2+\frac{b_iq_j^2}{q_i^2}
+\frac{b_jq_i^2}{q_j^2}\bigg]+\sum_{i=n-m+1}^nb_i\,,
\end{align*}
where $1<m\leq n$. Here $C^{(n)}=C_{(n)}=C$ is the
Casimir~\eqref{Casimir} and the functions
\[
\big\{H_0,C^{(l)},C_{(m)}:1<l<n,1<m\leq n\big\}
\]
are functionally independent. Moreover, the subsets
$\{H_0,C^{(m)}:1<m\leq n\}$ and $\{H_0,C_{(m)}:1<m\leq n\}$ are in
involution.
\end{theorem}

The previous result obviously applies to the Hamiltonian~\eqref{H},
which corresponds to
\[
H=\frac{J_+-c}{2(\ka+J_-)}\,.
\]
The main result of this paper is that the remaining first integral,
which makes $H$ MS, can indeed be found easily.

\begin{theorem}\label{T:main}
The remaining first integral can be chosen as
\begin{equation}\label{Ii}
I_i(\bp,\bq)=p_i^2-2\ms H(\bp,\bq)\,q_i^2+b_iq_i^{-2}\,,
\end{equation}
for any $1\leq i\leq n$. Moreover, the set $\{I_i:1\leq i\leq n\}$
is also in involution.
\end{theorem}
\begin{proof}
The equations of motion under $H$ are given by
\begin{align}
\dq_i&=\frac{p_i}{\ka+\bq^2}\,,\label{eq1}\\
\dot p_i&=\frac{2Hq_i+b_iq_i^{-3}}{\ka+\bq^2}\label{eq2}\,.
\end{align}
Combining both equations one can write
\begin{align*}
(\ka+\bq^2)\ddot
q_i+2(\bq\cdot\dot\bq)\ms\dq_i-\frac{2Hq_i+b_iq_i^{-3}}{(\ka+\bq^2)}=0\,.
\end{align*}
Multiplying this equation by $2(\ka+\bq^2)\ms\dq_i$ one immediately
finds
\[
\frac\dd{\dd t}\big[(\ka+\bq^2)^2\dq_i^2-2H\ms
q_i^2+b_iq_i^{-2}\big]=0\,,
\]
which yields the desired result. The functional independence of
$I_i$ is easily established through a tedious but straightforward
computation. The fact that $\{I_i\}$ are in involution follows from
 \begin{align*}
\{I_i,I_j\}&=
-2q_i^2\big(\{H,p_j^2\}-2H\,\{H,q_j^2\}+b_j\ms\{H,q_j^{-2}\}\big)
\\
&\hspace{16ex}+2q_j^2\big(\{H,p_i^2\}-2H\,\{H,q_i^2\}+b_i\ms\{H,q_i^{-2}\}\big)\\
&=-2q_i^2 \{H,I_j\}+2q_j^2 \{H,I_i\}=0\,.
\end{align*}
\end{proof}

Quadratic integrability is linked to the separability of the HJ
equation by a theorem due to Kalnins and Miller~\cite{KM84}. In
fact, it is not difficult to show that this equation is actually
superseparable. If we write $S(t,\bq)=W(\bq)-\frac12Et$ the HJ
equation for $H$ reads
\[
\bigg(\frac{\pd W}{\pd\bq}\bigg)^2-E\ms\bq^2+\sum_j
b_jq_j^{-2}=c+\ka E\,.
\]
Thus one essentially recovers the HJ equation for the SW system with
a different set of constants, namely
\[
H\sub{SW}(\bp,\bq)=\bp^2-E\bq^2+\sum_jb_jq_j^{-2}\,.
\]
Therefore,

\begin{proposition}
$H$ separates in the same coordinate systems as the SW system.
\end{proposition}

In three dimensions, e.g., it separates in 8 out of the 11 possible
orthogonal coordinate systems~\cite{Ka86,Ev91}. As a matter of fact,
it has long been conjectured that any MS system should separate in
multiple coordinate systems, but to our best knowledge this general claim
has not been proved or disproved.

\section{Integration of the equations of motion}
\label{S:eqs}

In this section we shall compute the trajectories of the Hamiltonian
system~\eqref{H} in closed form. It is convenient to start by
exploiting the $\fsl(2)$ symmetry to compute the evolution of the
radial variable
\[
x=\ka+J_-=\ka+\bq^2\,.
\]
We use the notation
\[
E:=2\ms H=\frac{J_+-c}x
\]
and assume $c\neq0$.

The evolution of the $\fsl(2)$ generators~\eqref{Jep} is given by
the following set of equations:
\begin{subequations}
\begin{align}
\dot x&=\{J_-,H\}=\frac{2J_0}x\,,\label{dx}\\
\dot J_0&=\{J_0,H\}=\frac{J_++E(x-\ka)}x\,,\label{dJ_0}\\
\dot J_+&=\{J_+,H\}=\frac{2EJ_0}x\,.\label{dJ_+}
\end{align}
\end{subequations}
Using these equations one can write the Casimir~\eqref{Casimir} as
\begin{align*}
C&=-\tfrac14x^2\dot x^2+Ex^2+(c-E\ka)x-c\ka\geq0\,.
\end{align*}
For the sake of concreteness, let us assume that $E>0$ and set
\begin{align}
\al&:=\frac12\Big(\ka-\frac cE\Big)\,\,,\label{aldef}\\
\ga^2&:=\frac14\Big(\ka+\frac cE\Big)^2+\frac CE\,.\label{ga}
\end{align}
In this case
\begin{equation}\label{Eq.x}
x^2\dot x^2=4E\big[(x-\al)^2-\ga^2\big]\,,
\end{equation}
and one can easily integrate this equation as
\begin{equation}\label{x}
\pm2{\sqrt{E}}(t-\tau)=\sqrt{(x-\al)^2-\ga^2}+
\al\,\cosh^{-1}\Big(\frac{x-\al}{\ga}\Big)\,.
\end{equation}
Here $\tau$ is an arbitrary constant and $\ga$ is the positive
square root of $\ga^2$. Note that the equation~\eqref{Eq.x} for the
variable $x$ (the squared radius) coincides with the radial equation
in the Kepler problem.

The above expression yields $t$ as a monotonic function of $x$, so this
relation is globally invertible (in each half-orbit). However, it is not
possible to obtain the inverse function $x(t)$ in closed form. Hence we prefer
to parametrize the trajectory by the radial variable $x$. In terms of this
variable, the first integral~\eqref{Ii} reads
\begin{equation}\label{Iinew}
I_i=4E\big[(x-\al)^2-\ga^2\big]\Big(\frac{\dd q_i}{\dd
x}\Big)^2-Eq_i^2+b_iq_i^{-2}\,.
\end{equation}
Let us set
\begin{equation}\label{vars}
Q_i=q_i^2,\qquad \al_i=-\frac{I_i}{2E},\qquad \ga^2_i=\al_i^2+E^{-1}b_i,
\end{equation}
so that
\[
Eq_i^4+I_iq_i^2-b_i=E\big[(Q_i-\al_i)^2-\ga^2_i\big]\,.
\]
Then one can perform the integration of Eq.~\eqref{Iinew} to obtain
\[
\int\frac{\dd
Q_i}{\sqrt{(Q_i-\al_i)^2-\ga^2_i}}={\vp_i}
+\int\frac{\dd x}{\sqrt{(x-\al)^2-\ga^2}}\,,
\]
i.e.,
\begin{align*}
\cosh^{-1}\bigg(\frac{Q_i-\al_i}{\ga_i}\bigg)&=\vp_i+\cosh^{-1}\Big(\frac{x-\al}{\ga}\Big)\,,
\end{align*}
where $\vp_i$ are constants and $\ga_i\ge0$. The full solution of
this equation is therefore given by
\begin{align}\label{Qi}
Q_i&=\al_i+\ga_i\,\cosh(X+\vp_i)\notag\\
&=\al_i+\ga^{-1}\ga_i\,\cosh\vp_i\,(x-\al)+\ga_i\,\sinh\vp_i\bigg|1-\Big(\frac{x-\al}\ga\Big)^2\bigg|^{1/2}\,,
\end{align}
with $X:=\cosh^{-1}(\ga^{-1}(x-\al))$. Eqs.~\eqref{x} and~\eqref{Qi}
are the main results of this section.

All the orbits with positive energy are recovered through an
appropriate choice of the $2n$ parameters
\begin{equation}\label{fundamental}
\big\{\tau,\al_i,\vp_j:1\leq i\leq n-1,\,1\leq j\leq n\big\}\,,
\end{equation}
where $\tau$ was defined in Eq.~\eqref{x}. In what follows we shall
express the remaining quantities in terms of this fundamental set.

If we sum over $i$ in Eq.~\eqref{Qi} and use that $x=\ka+\sum_i
Q_i$, we immediately find
\[
x-\ka=\sum_i\al_i+\ga^{-1}(x-\al)\sum_i\ga_i\,\cosh\vp_i+
\bigg|1-\Big(\frac{x-\al}\ga\Big)^2\bigg|^{1/2}\sum_i\ga_i\,\sinh\vp_i\,.
\]
Therefore we reach the following compatibility conditions:
\begin{subequations}
\begin{align}
\sum_i\al_i+\ka&=\al\,,\label{al}\\
\sum_i\ga_i\,\cosh\vp_i&=\ga\,,\label{be}\\
\sum_i\ga_i\,\sinh\vp_i&=0\,.\label{constr}
\end{align}
\end{subequations}
Eq.~\eqref{al} gives the value of $\al$ in terms of $\{\al_j\}$ and
Eq.~\eqref{be} consistently provides the value of the Casimir by
means Eq.~\eqref{ga}. Eq.~\eqref{constr}, which imposes the
constraint
\[
\sum_i\sinh\vp_i\,\big(\al_i^2+E^{-1}b_i\big)^{1/2}=0
\]
on the the parameters $\{\al_i,\vp_j\}$ by virtue of Eq.~\eqref{vars}, is
responsible for the fact that only $2n-1$ among these parameters have been
included in the set~\eqref{fundamental}. Note that Eqs.~\eqref{aldef}
and~\eqref{al} can be combined to express $E$ as a function of $\{\al_i\}$ as
\[
E=-\frac c{\ka+2\sum_i\al_i}\,.
\]
Finally, this permits to express $\ga_i$ in terms of $\{\al_j\}$ by
means of Eq.~\eqref{vars}. We omit the discussion of the case
$E\leq0$, which goes along the same lines.

\section{The geometric content of $\cH$}
\label{S:geometry}

It is apparent that $\cH$ is a (nonconstant) multiple of the SW
system. What we want to stress in this section is that their
connection does not end here: $\cH$ is in fact the \emph{intrinsic}
SW Hamiltonian in the manifold $\cM^n$. And what makes this space remarkable is
that its intrinsic SW model is also MS.

It is apparent from Eq.~\eqref{ds2} that the Riemannian manifold
$\cM^n$ is spherically symmetric, this $\SO(n)$ symmetry being a
consequence of the $\fsl(2)$ coalgebra structure outlined in
Sec.~\ref{S:int}. Its curvature is certainly nonconstant; for the
sake of completeness we note that its scalar curvature is negative
and given by
\[
R=-(n-1)\frac{3(n-2)\ms\bq^2+2\ka n}{(\ka+\bq^2)^3}\,.
\]

One can define intrinsic versions of the Kepler and harmonic
oscillator potentials on $\cM^n$. To this end, denote by
$\Delta_{\cM^3}$ the Laplace--Beltrami operator on $\cM^3$ and let
$V_3\in C^\infty(\cM^3\backslash\{\mathbf 0\})$ be its (minimal
symmetric) Green function, i.e., a function $V_3(\bq)=v(|\bq|)$
satisfying
\[
-\Delta_{\cM^3}V_3=\delta_{\cM^3}\,.
\]
Here $v\in C^\infty(\RR^+\backslash\{0\})$ and $\delta_{\cM^3}$
stands for the delta distribution in $\cM^3$ supported at the
origin. It can be proved~\cite{EP06d} that such a function exists
and is unique.

\begin{definition}
The \emph{Kepler and harmonic oscillator potentials} in $\cM^n$ are
\begin{align*}
V\Kep(\bq):=K\,v(|\bq|)\,,\qquad V\Harm(\bq):=K\,v(|\bq|)^{-2}\,,
\end{align*}
where $K\in\RR$ is an arbitrary constant.
\end{definition}
\begin{remark}
The above definition, which is based on the potential theory of
these spaces, can be easily extended to any spherically symmetric
Riemannian space with the appropriate behavior at
infinity~\cite{LT87}. Such a prescription reproduces the intrinsic
Kepler and harmonic potentials on constant curvature spaces
(see~\cite{CRS05} and references therein). The case of
$\U(2)$-symmetric K\"ahler 4-manifolds, which had already appeared
in the literature~\cite{NY04}, is based on the same ideas but does
not fit into this framework. Certainly the above prescription does
not generally lead to MS potentials.
\end{remark}

Let $\bq\in\cM^3$ and define $r=|\bq|$. It can be readily verified
that the action of the Laplace--Beltrami operator in $\cM^3$ on some
function $f(r)$ is
\[
\Delta_{\cM^3}f=\frac1{r^2(\ka+r^2)}\frac \dd{\dd
r}\bigg(r^2\sqrt{\ka+r^2}\,\frac{\dd f}{\dd r}\bigg)\,.
\]
By setting the above expression equal to zero for $r>0$ it is
straightforward to find that the function $v$ defined above is
\[
v(r)=\frac{\sqrt{\ka+r^2}}r
\]
up to a multiplicative constant. Hence the Kepler and harmonic
oscillator potentials in $\cM^n$ are respectively given by
\[
V\Kep(\bq)=K\ms\frac{\sqrt{\ka+\bq^2}}{|\bq|}\,,\qquad
V\Harm(\bq)=\frac{K\ms\bq^2}{\ka+\bq^2}\,.
\]
This shows that the MS Hamiltonian~\eqref{cH} is in fact the
intrinsic SW systems, i.e.,
\[
\cH(\bp,\bq)=\|\bp\|^2_{\cM^n}+V\Harm(\bq)+(\ka+\bq^2)^{-1}\sum_jb_j\ms
q_j^{-2}\,,
\]
with $\|\cdot\|_{\cM^n}$ representing the metric on the cotangent
bundle of $\cM^n$.

\section*{Acknowledgements}

This work was partially supported by the Spanish MEC and the Junta
de Castilla y Le\'on under grants no.\ FIS2004-07913 and VA013C05
(A.B.\ and F.J.H.), by the Spanish DGI under grant no.\
FIS2005-00752 (A.E.) and by the INFN--CICyT (O.R.). Furthermore,
A.E. acknowledges the financial support of the Spanish MEC through
an FPU scholarship, as well as the hospitality and the partial support of the Physics Department of Roma Tre University.


\end{document}